\documentclass[graybox]{svmult}

\usepackage{mathptmx}       % selects Times Roman as basic font
\usepackage{helvet}         % selects Helvetica as sans-serif font
\usepackage{courier}        % selects Courier as typewriter font
\usepackage{type1cm}        % activate if the above 3 fonts are 
\usepackage{makeidx}         % allows index generation
\usepackage{graphicx}        % standard LaTeX graphics tool
\usepackage{multicol}        % used for the two-column index
\usepackage[bottom]{footmisc}% places footnotes at page bottom

\makeindex             % used for the subject index
\graphicspath{{converted_graphics/}}
\begin{document}

\title*{On the problem of synchronization of identical dynamical systems:
The Huygens's clocks}
\titlerunning{Synchronization of Huygens's clocks}
% for an abbreviated version of
\author{Rui Dil\~ao}
\institute{Rui Dil\~ao \at NonLinear Dynamics Group, Instituto Superior T\'ecnico
Av. Rovisco Pais, 1049-001 Lisbon, Portugal,
\email{rui@sd.ist.utl.pt}}
\maketitle

\abstract{In 1665, Christiaan Huygens reported the observation of the synchronization of two pendulum clocks hanged on the wall of his workshop. After synchronization, the clocks swung exactly in the same frequency and $180^{o}$ out of phase --- anti-phase synchronization.  Here, we propose and analyze a new interaction mechanism between oscillators leading to exact anti-phase and in-phase synchronization of pendulum clocks, and we  determine a sufficient condition for the existence of an exact anti-phase synchronization state. We show that exact anti-phase and in-phase synchronization states can coexist in phase space, and the periods of the synchronized states are different from the eigen-periods of the individual oscillators.  We  analyze the robustness of the system when the parameters of the individual pendulum clocks are varied, and we show numerically that exact anti-phase and in-phase synchronization states  exist in systems of coupled oscillators with different parameters.}

\abstract*{In 1665, Christiaan Huygens reported the observation of the synchronization of two pendulum clocks hanged on the wall of his workshop. After synchronization, the clocks swung exactly in the same frequency and $180^{o}$ out of phase --- anti-phase synchronization.  Here, we propose and analyze a new interaction mechanism between oscillators leading to exact anti-phase and in-phase synchronization of pendulum clocks, and we  determine a sufficient condition for the existence of an exact anti-phase synchronization state. We show that exact anti-phase and in-phase synchronization states can coexist in phase space, and the periods of the synchronized states are different from the eigen-periods of the individual oscillators.  We  analyze the robustness of the system when the parameters of the individual pendulum clocks are varied. Moreover, we show that exact anti-phase and in-phase synchronization states can exist in systems of coupled oscillators with different parameters.} 

\section{Introduction}
\label{sec:1}

In 26 February 1665, Christiaan Huygens, in a letter to his father,
\cite{1}, reported the observation of the synchronization of two pendulum
clocks closely hanged on the wall of his workshop. After
synchronization, the clocks swung exactly in the same frequency and
180$^{o}$ out of phase. For attachment distances, less than 1 meter,
the clocks always synchronize with the 180$^{o}$ phase difference. For
larger attachment distances ($>4.5$~m),
synchronization, if it occurs, takes longer times. Huygens also noted
that if the two clocks were hanged in such a way that the planes of
oscillation of the two pendulums were perpendicular, then 
synchronization did not occur.  The Huygens observations were the
first time that synchronization effects have been described
scientifically.

Huygens justified the observed synchronization phenomena by the
``sympathy that cannot be caused by anything other
than the imperceptible stirring of the air due to the motion of the
pendulum''.

In recent years, there has been a growing interest in the detailed analysis of synchronization phenomena, both from the theoretical and the experimental points of views. From the experimental point of view, 
Bennett {\it et al.}, \cite{2}, built an experimental device   consisting of two interacting pendulum clocks hanged on a heavy support, and  this support was mounted on a low-friction wheeled cart. This device  moves by the action of the tensions due to the swing of the two pendulums, and the interaction between the two clocks is caused by the mobility of the heavy base of the clocks. 
With this device, the anti-phase synchronization mode is reached when the difference between the natural or eigen-frequencies of the two clocks is less than $0.0009$~Hz. If the difference between these frequencies is larger than $0.0045$~Hz, the two clocks do not  synchronize, running ``uncoupled'' or in a state of beating death, \cite{2}. This situation is unsatisfactory when compared with the observations of Huygens. For example, a difference of order of $\Delta \omega=0.0009$~Hz for the two pendulum eigen-frequencies  corresponds to a difference in the lengths of the pendulum rods of the order of  $\Delta \ell = \sqrt{g}\ell^{3/2}\Delta \omega /4\pi$, which gives, for $\ell =1$~m and $g=9.8$~ms$^{-2}$,  $\Delta \ell=4$~mm, and  for  $\ell =0.178$~m (the length of the pendulum rods used by Huygens, \cite{1}), $\Delta \ell=0.02$~mm, a precision that Huygens certainly could not achieve. 
According to Bennett {\it et al.} \cite[p. 578]{2},   Huygens's results depended on both talent and luck.

In the experiment of Bennett {\it et al.}, \cite{2}, the in-phase synchronization is the natural way of synchronization
of the two pendulum clocks. However, due to a detailed description of the Huygens
findings, we can believe that the in-phase synchronization was never
 observed by Huygens.

Another experimental model mimicking the Huygens's clocks system,
consists of two pendulums whose suspension rods are connected by a weak string, and one of the two pendulums is driven by an external rotor, \cite{3} and \cite{4}.
 In this system, the in-phase synchronization is approximately achieved with a small phase shift, and  the experimental measurements and the model analysis both agree. 
The numerical results of Fradkov and Andrievsky for this device, \cite{4}, show simultaneous and  approximate in-phase and anti-phase synchronization, tuned by different initial conditions.
In another experimental device made of two rotors controlled by external torques  (\cite{5, 6}), Andrievsky {\it et al.}, \cite{5},
reported approximate anti-phase and in-phase synchronization of the
two  oscillators. In this experiment, the synchronization  parameter is the stiffness of a string connecting the two rotors. 

For demonstration purposes, Pantaleone, \cite{pan}, reported an experimental device constructed with two metronomes on a freely moving base. In this metronomes experiment, the phase difference of the synchronous state is close to  $0^{o}$.
Increasing the damping effect on the freely moving base, the author also reported approximate  synchronization with a difference in phase close to $180^{o}$. 
From the theoretical point of view, the  equations describing this experimental device lead to the Kuramoto synchronization model, \cite{kura}, where the synchronization mechanism is associated with a non-linear effect associated with the  
phase difference between the oscillators. 

In the Bennett {\it et al.}, \cite{2}, and the Pantaleone, \cite{pan}, experimental systems, the interaction mechanism  between oscillators is obtained by a moving base, an idea advanced by Kortweg in 1906, \cite{kort}. 

In the  experimental systems described above, there is no clear evidence
of what mechanisms are in the origin of  the anti-phase synchronization, as described by Huygens. In some experiments, it appears that if the pendulums have slightly different periods, the two oscillators may not synchronize,
\cite{2, 4}. 

However, as this special type of collective rhythmicity occurs in biological systems and several other natural phenomena, \cite{7}, where individual  periods are different, it is important to derive and to understand the interaction mechanisms leading to exact synchrony. 
Besides all the attention of the scientific community for this
synchronization phenomenon, there is no clear evidence of a mechanism,
leading to anti-phase synchronization.

In this paper, we propose and analyze a new interaction mechanism between oscillators leading to exact anti-phase and in-phase synchronization. The synchronization mechanism is obtained with  a  damped  elastic string, and the
oscillators under analysis can be simple harmonic oscillators,
pendulums, and any type of non-linear oscillators with a limit cycle in phase space, as is the case of pendulum clocks, \cite{8}.  
The main result of this paper is to show that exact anti-phase synchronization can  always be achieved for systems of coupled oscillators.   

This paper is organized as follows. In section~\ref{sec:2}, we introduce the synchronization  model in its full generality,  discussing its physical assumptions, and we derive the equations of motion of the interacting oscillators. Then, we make the approximation of small oscillations. In section~\ref{sec:3}, we introduce a simplified pendulum clock model, and we prove that the ordinary differential equation describing the dynamics of the clock has a limit cycle in phase space. Based on the linear model derived in section~\ref{sec:2}, in section~\ref{sec:4}, we discuss the concept of anti-phase synchronization, and we find a sufficient condition for the existence of an exact anti-phase synchronized state for the two pendulum clock system. 
This result is stated as Theorem~\ref{t1}, the main result of this paper. After exploring numerically the phase space structure of the solutions of the model equations as a function of a control parameter,
we show the existence of exact anti-phase and in-phase synchronization states for the Huygens's two pendulum clocks system.  
The tuning between the two types of synchronization regimes can be controlled through a damping constant associated with the (elastic) interaction mechanism and by the choice of initial conditions.
These results justify some numerical results published previously, \cite{9}. 
In section~\ref{sec:5}, we analyze numerically the persistence of in-phase and anti-phase synchronization phenomena when we vary the parameters of the individual pendulum clocks. We show that the anti-phase and the in-phase synchronization regimes persist even if the two oscillators have different eigen-periods and parameters.
Finally, in  section~\ref{sec:6}, we resume the conclusions of the paper.

\section{A model for the synchronization of the two pendulum clocks}
\label{sec:2}

\begin{figure}
\begin{center}
\includegraphics[width=10.0cm,height=3.45cm]{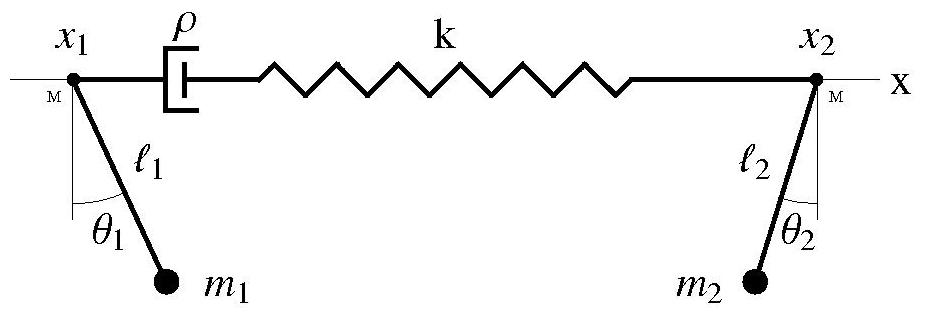}
\end{center}
 \caption{Model to analyze the synchronization of the Huygens's two-pendulum clocks system. The two pendulums are a representation of  two nonlinear oscillators. The interaction between the pendulums is done by the tension forces  at the attachment points, and they actuate through an elastic and resistive media. Each attachment points is considered to have  mass $M$.}
\label{fig1} 
\end{figure}

In the Huygens two pendulum clocks system,
the pendulums are hanged in a common support, and  the only possible interaction is due to the tension forces generated by the oscillatory motion of the two pendulums. These tension forces propagate  through the common support, that we consider to be elastic.  
The role of the tension forces in the synchronization mechanism
is corroborated by the Huygens's finding that when 
the planes of oscillation of the two hanged pendulums are mutually perpendicular,
no synchronization is observed. In fact, the components of the tension forces generated by the motion of the pendulums are in the plane of motion of the pendulums, and force the motion of the two attachment points.  To be more
specific, we consider the geometric arrangement of Fig.~\ref{fig1}, where the
two pendulums have masses $m_1$ and $m_2$, and lengths 
$\ell_1$ and $\ell_2$, respectively.

The pendulums are considered connected by a massless string with
stiffness constant $k$. The perturbations that
propagate along the string are damped, and the damping force is
proportional to the   velocity of the attachment points of the string, with damping constant
$\rho $. The damped string simulates the interaction effects that propagate through
the elastic and resistive media. We consider that the attachment points of the
pendulums, located at the horizontal coordinates $x = x_1
$ and $x = x_2 $, have equal masses, and we
denote this mass by $M$. As we shall see, the
introduction of this mass is necessary to obtain explicitly the
equations of motion.

The system of Fig.~\ref{fig1}, considered without the damping forces, is
described by the four degrees of freedom Lagrangian,
\begin{equation}
\begin{array}{ll}\displaystyle
 L  =&\frac{1}{2}m_1 (\ell_1^2 \dot \theta_1^2  + \dot x_1^2  + 2\ell_1 \dot x_1 \dot \theta_1 \cos \theta_1 )+ m_1 g\ell_1 \cos \theta_1   \\ [2pt]
&+ \frac{1}{2}m_2 (\ell_2^2 \dot \theta_2^2  + \dot x_2^2  + 2\ell_2 \dot x_2 \dot \theta_2 \cos \theta_2 ) + m_2 g\ell_2 \cos \theta_2\\ [2pt]
&+ \frac{1}{2}M(\dot x_1^2  + \dot x_2^2 ) - \frac{1}{2}k(x_2  - x_1 )^2 \, ,
 \end{array}
\label{eq1}
\end{equation}
where  $\theta_1$ and $\theta_2$ are the angular coordinates of the two pendulums,   $g$ is the acceleration due to the gravity force, and the last two terms describe the  interaction mechanism between the two pendulums.  From (\ref{eq1}), the Lagrange equations of motion of the two interacting pendulums are,
\begin{equation}
\begin{array}{l}
m_1 \ell_1 \ddot \theta_1  + m_1 g\sin \theta_1  =  - m_1 \ddot
x_1 \cos \theta_1  \\[2pt]
m_2 \ell_2 \ddot \theta_2  + m_2 g\sin \theta_2  =  - m_2 \ddot
x_2 \cos \theta_2  \\[2pt]
(M + m_1 )\ddot x_1  + m_1 \ell_1 \ddot \theta_1 \cos \theta_1  =
m_1 \ell_1 \dot \theta_1^2 \sin \theta_1  + k(x_2  - x_1 ) \\[2pt]
(M + m_2 )\ddot x_2  + m_2 \ell_2 \ddot \theta_2 \cos \theta_2  =
m_2 \ell_2 \dot \theta_2^2 \sin \theta_2  - k(x_2  - x_1 ) \, .
\end{array}
\label{eq2}
\end{equation}
Introducing the damping effects into system of equations (\ref{eq2}), we obtain,
\begin{equation}
\begin{array}{l}
m_1 \ell_1 \ddot \theta_1  + f_1 (\theta_1 ,\dot \theta_1 ) + m_1
g\sin \theta_1  =  - m_1 \ddot x_1 \cos \theta_1  \\[2pt]
m_2 \ell_2 \ddot \theta_2  + f_2 (\theta_2 ,\dot \theta_2 ) + m_2
g\sin \theta_2  =  - m_2 \ddot x_2 \cos \theta_2  \\[2pt]
(M + m_1 )\ddot x_1  + 2\rho \dot x_1  + m_1 \ell_1 \ddot \theta_1
\cos \theta_1  = m_1 \ell_1 \dot \theta_1^2 \sin \theta_1  + k(x_2 
- x_1 ) \\[2pt]
(M + m_2 )\ddot x_2  + 2\rho \dot x_2  + m_2 \ell_2 \ddot \theta_2
\cos \theta_2  = m_2 \ell_2 \dot \theta_2^2 \sin \theta_2  - k(x_2 
- x_1 )\, ,
\end{array}
\label{eq3}
\end{equation}
where $\rho$  is the damping constant of the attachment points of the pendulums, and the functions  $f_1 (\theta_1 ,\dot \theta_1 )$ and $f_2 (\theta_2 ,\dot \theta_2 )$  describe the escaping mechanism
of the pendulum clocks  (\S~\ref{sec:3}).

The system of equations (\ref{eq3}) implicitly defines a system of ordinary differential
equations. If $M > 0$, $\ell_1  >
0$, $m_1  > 0$, $\ell_2  >
0$ and $m_2  > 0$, the system (\ref{eq3}) can be solved
algebraically in order to the higher derivatives, and we obtain,
\begin{equation}
\begin{array}{l}
m_1 \ell_1 \ddot \theta_1  + f_1 (\theta_1 ,\dot \theta_1 ) + m_1
g\sin \theta_1  =  - m_1 \cos \theta_1 F_1  \\[2pt]
m_2 \ell_2 \ddot \theta_2  + f_2 (\theta_2 ,\dot \theta_2 ) + m_2
g\sin \theta_2  =  - m_2 \cos \theta_2 F_2  \\
\ddot x_1  = F_1  \\[2pt]
\ddot x_2  = F_2 \, , 
\end{array}
\label{eq4}
\end{equation}
where,
\begin{equation}
\begin{array}{l}\displaystyle
F_1  = \frac{{f_1 (\theta_1 ,\dot \theta_1 )\cos \theta_1  + m_1
g\sin \theta_1 \cos \theta_1  - 2\rho \dot x_1  + m_1 \ell_1 \dot
\theta_1^2 \sin \theta_1  + k(x_2  - x_1 )}}{{M + m_1 \sin ^2 \theta
_1 }} \\ \displaystyle  
F_2  = \frac{{f_2 (\theta_2 ,\dot \theta_2 )\cos \theta_2  + m_2
g\sin \theta_2 \cos \theta_2  - 2\rho \dot x_2  + m_2 \ell_2 \dot
\theta_2^2 \sin \theta_2  - k(x_2  - x_1 )}}{{M + m_2 \sin ^2 \theta
_2 }}\, .
\end{array}
\label{eq5}
\end{equation}

The system of ordinary differential
equations (\ref{eq5}) are a synchronization model 
for the Huygens's two pendulum clocks system, \cite{9}.
Here, we will consider only the case $M > 0$. 
The case $M =0$  will be analyzed elsewhere.

For  small amplitude of oscillations in $\theta_1$ and $\theta_2$, 
the  system of equations (\ref{eq4})-(\ref{eq5}) simplify, and we obtain,
\begin{equation}
\begin{array}{l}\displaystyle 
m_1 \ell_1 \ddot \theta_1  + f_1 (\theta_1 ,\dot \theta_1 ) + m_1
g\theta_1  =  - \frac{m_1}{M} \left({f_1 (\theta_1 ,\dot \theta_1 )  + m_1
g \theta_1   - 2\rho \dot x_1   + k(x_2  - x_1 )} \right) \\[6pt] \displaystyle 
m_2 \ell_2 \ddot \theta_2  + f_2 (\theta_2 ,\dot \theta_2 ) + m_2
g\theta_2  =  - \frac{m_2}{M} \left( {f_2 (\theta_2 ,\dot \theta_2 )  + m_2
g \theta_2   - 2\rho \dot x_2   - k(x_2  - x_1 )}  \right)\\[6pt] \displaystyle 
\ddot x_1  = \frac{1}{M} \left({f_1 (\theta_1 ,\dot \theta_1 )  + m_1
g \theta_1   - 2\rho \dot x_1   + k(x_2  - x_1 )} \right)  \\[6pt] \displaystyle   
\ddot x_2  = \frac{1}{M} \left( {f_2 (\theta_2 ,\dot \theta_2 )  + m_2
g \theta_2   - 2\rho \dot x_2   - k(x_2  - x_1 )}  \right)   \, .
\end{array}
\label{eq6}
\end{equation}
In this paper, our goal is to analyze the synchronization 
properties of the solutions of the system of ordinary differential equations (\ref{eq6}).

To model the Huygens's two pendulum clocks experiment,  we have to  choose a specific form for the functions $f_1 (\theta_1 ,\dot \theta_1 )$ and $f_2 (\theta_2 ,\dot \theta_2 )$, describing  the oscillatory behavior of pendulum clocks.

\section{A simple clock model}
\label{sec:3}

To restore the energy lost by a pendulum clock during one period, the sustained oscillations  can be maintained by an impulsive force acting on the pendulum rod,
or by the force originated by the smooth unwinding of a  circular string attached to the pendulum balance wheel, \cite[pp. 169-200]{8}. In any case, the dynamics of a pendulum clock is modeled by a  
two-dimensional dynamical system with a limit cycle in phase space. This same qualitative behavior can be obtained  with an oscillator actuated by a non-linear damping force, piecewise proportional to the angular velocity of the pendulum. For small amplitude of oscillations, the  proportionality constant is positive, and,  for large amplitudes of oscillations, the  proportionality constant is negative. 

To simplify our analysis, we take, as a qualitative model for a pendulum clock, the following second order differential
equation,
\begin{equation}
m\ell
\ddot \theta  + f(\theta; \lambda ,{\tilde \theta})\dot \theta  + m g \theta  = 0\, ,
\label{eq3:1}
\end{equation}
where,
\begin{equation}
f(\theta; \lambda ,{\tilde \theta}) = \left\{ 
   \begin{array}{rcl}
-2\lambda   &\ \hbox{if}\ & |\theta | < \tilde \theta  \\ 
2\lambda     &\hbox{if}\ & |\theta | \ge \tilde \theta \, .
\end{array} 
\right.  
\label{eq3:2}
\end{equation}
The function  $-f(\theta; \lambda ,{\tilde \theta})\dot \theta$ is the damping force of the pendulum clock, and $\lambda$ and ${\tilde \theta}$ are positive constants.  

\begin{proposition}\label{p1} If $\lambda >0$, $\tilde \theta >0$, $m>0$, $\ell >0$, and $g>0$, the second order differential equation
(\ref{eq3:1}), with the damping function (\ref{eq3:2}), has a unique and stable limit cycle in phase space.
\end{proposition}

\begin{proof} Assume that $\lambda >0$, $\tilde \theta >0$, $m>0$, $\ell >0$, and $g>0$. Define the function,
\[
F(\theta; \eta,\tilde \theta)=\frac{1}{m\ell} \int_0^{\theta} f(s; \lambda,\tilde \theta)ds=\left\{
   \begin{array}{llcl}
-&2\eta \theta  &\ \hbox{if}\ & |\theta | < \tilde \theta  \\ 
&2\eta  \theta- 4\eta  \tilde \theta  &\hbox{if}\ & |\theta | \ge \tilde \theta \, ,
\end{array} \right.
\]
where, $\eta=\lambda/(m\ell)$, and $\omega^2=g/\ell$. 
With the new coordinate, $x=\dot \theta+F(\theta; \eta,\tilde \theta)$, the differential equation (\ref{eq3:1}), with damping function (\ref{eq3:2}), has the 
Li{\'e}nard form,
\begin{equation}
\left\{
   \begin{array}{lcl}
\dot \theta &=& x-F(\theta; \eta,\tilde \theta)\\  
\dot x &=& -\omega^2 \theta \, .
\end{array} \right.  
\label{eq3:3}
\end{equation}
where  $F(\theta; \eta,\tilde \theta)$ is a continuous and odd function of $\theta$. The existence and stability 
of a limit cycle solution for equation (\ref{eq3:3})  follows from the 
Li{\'e}nard theorem, \cite[pp. 179-181]{Hart}.
\qed
\end{proof}

\begin{figure}
\begin{center}
\includegraphics[width=10.0cm,height=4.94cm]{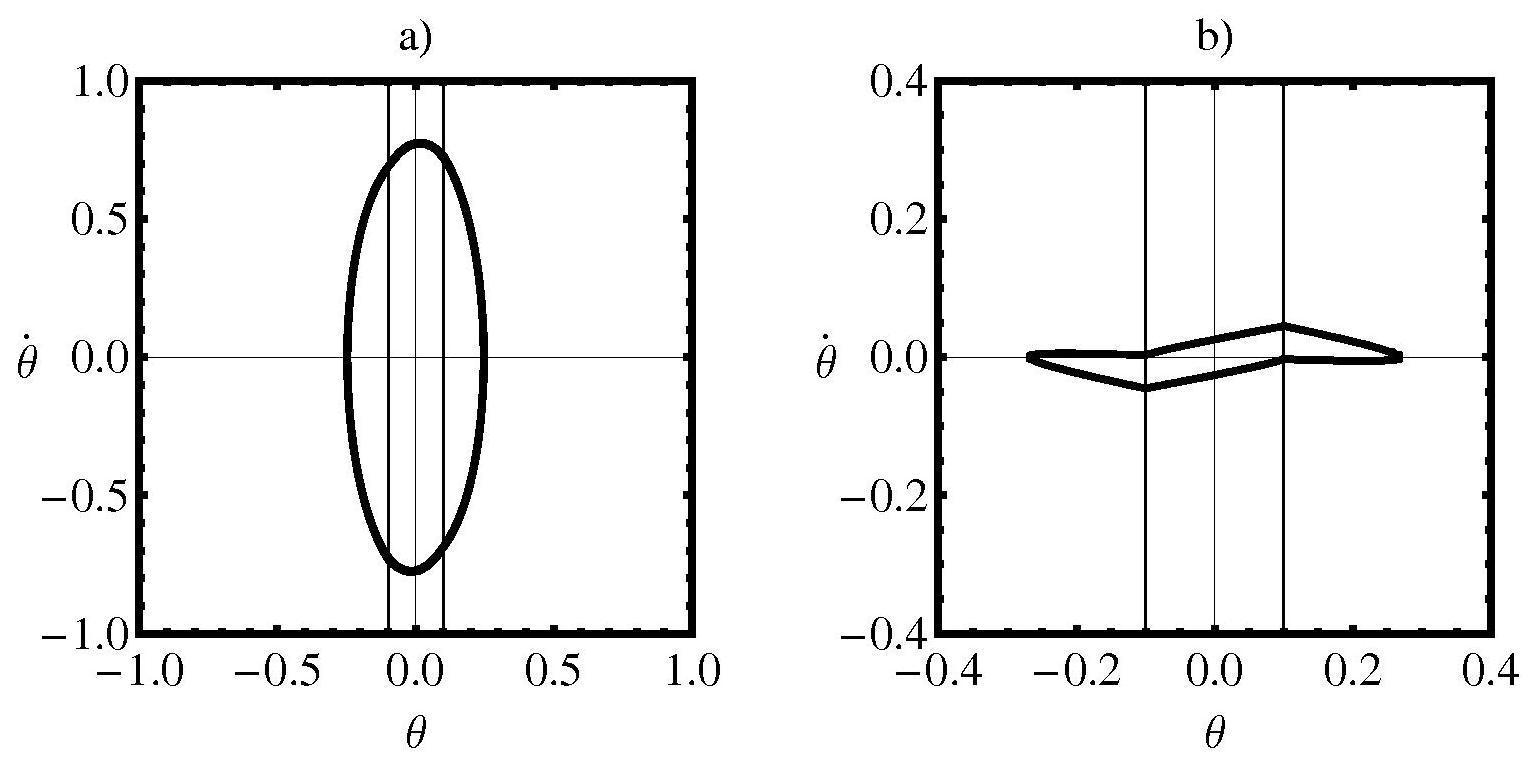}
\end{center}
 \caption{Limit cycles solutions of the differential  equation (\ref{eq3:1}), with the damping function (\ref{eq3:2}).
In a) we have chosen $\omega^2=9.8$, and in b), $\omega^2=0.005$, where $\omega^2=g/\ell$, $m=1$ and $\ell=1$.
The parameter values of the damping function are, $\lambda  =
0.1$ and  $\tilde \theta  = 0.1$. The two vertical lines for $\theta=\pm \tilde \theta$ show the  discontinuity of the tangential derivative along the limit cycles.}
\label{fig2} 
\end{figure}

In the conditions of Proposition~\ref{p1}, the differential equation (\ref{eq3:1}), with   damping function (\ref{eq3:2}), has a unique limit cycle in phase space. In
Fig.~\ref{fig2}, we show the limit cycle in phase space  of this non-linear
oscillator, for two values of the parameter $\omega^2=g/\ell$.

The non-linear oscillator model defined by the  differential equation (\ref{eq3:1}), with damping function (\ref{eq3:2}), has all the qualitative properties found in  more accurate pendulum clock models, 
\cite{8}, and will be used in the next sections to model Huygens's clocks.

\section{Synchronization of two pendulum clocks  with equal parameters}
\label{sec:4}

By (\ref{eq6}), the two interacting pendulum clocks with equal parameters are modeled the system of  differential equations (\ref{eq6}),
\begin{equation}
\begin{array}{l}\displaystyle
\ddot \theta_1  + (\frac{1}{m\ell}+\frac{1}{M\ell})f (\theta_1 ;\lambda ,\tilde \theta )\dot \theta_1 + \omega^2(1+\frac{m}{M})\theta_1  -2\frac{\rho}{M\ell}\dot x_1=  - \frac{k}{M\ell} (x_2  - x_1 )  \\[6pt] \displaystyle
\ddot \theta_2  + (\frac{1}{m\ell}+\frac{1}{M\ell})f (\theta_2 ;\lambda ,\tilde \theta )\dot \theta_2 + \omega^2(1+\frac{m}{M})\theta_2  -2\frac{\rho}{M\ell}\dot x_2=   \frac{k}{M\ell} (x_2  - x_1 )\\[6pt] \displaystyle
\ddot x_1  - \frac{1}{M}  f (\theta_1 ;\lambda ,\tilde \theta )\dot \theta_1 -\frac{m}{M} 
g \theta_1   + 2\frac{\rho}{M} \dot x_1   = \frac{k}{M}(x_2  - x_1 )   \\[6pt] \displaystyle
\ddot x_2 - \frac{1}{M}  f (\theta_2 ;\lambda ,\tilde \theta )\dot \theta_2 -\frac{m}{M} 
g \theta_2  + 2\frac{\rho}{M} \dot x_2   = -\frac{k}{M}(x_2  - x_1 )\, ,    
\end{array}
\label{eq4:1}
\end{equation}
where $f(\theta; \lambda ,{\tilde \theta})$ is given by (\ref{eq3:2}), and $\omega^2=g/\ell$.

Our goals here is to show that the solutions of the system of differential equations 
(\ref{eq4:1})  synchronize in anti-phase. For that, we
begin by analyzing the   solutions of  the system of equations (\ref{eq4:1}) for the particular case where the initial conditions in $\theta_1$ and $\theta_2$ are bounded  by ${\tilde \theta}$, that is, $|\theta_1(0)|<{\tilde \theta}$  and $|\theta_2(0)|<{\tilde \theta}$. 
As the system of equations (\ref{eq4:1}), together with (\ref{eq3:2}), is piecewise linear, and if $|\theta_1(t)|<{\tilde \theta}$,  and $|\theta_2(t)|<{\tilde \theta}$,
for every $t\in [0,t^*]$, where $t^*$ is a positive constant,
then the piecewise linear system of equations (\ref{eq4:1}) can be written as the linear first order system of differential equations,
\begin{equation}
\left( {\begin{array}{l}
 \dot \theta_1    \\
 \dot v_1    \\
 \dot \theta_2 \\  
 \dot v_2  \\ 
 \dot x_1  \\ 
 \dot w_1  \\ 
 \dot x_2  \\ 
 \dot w_2  
\end{array}} \right) = 
\left( {\begin{array}{*{20}c}
   0 & 1 & 0 & 0 & 0 & 0 & 0 & 0  \\
    A & B & 0 & 0 & {\frac{k}{{M\ell }}} & {\frac{{2\rho }}{{M\ell }}} & { - \frac{k}{{M\ell }}} & 0  \\
   0 & 0 & 0 & 1 & 0 & 0 & 0 & 0  \\
   0 & 0 & A & B & { - \frac{k}{{M\ell }}} & 0 & {\frac{k}{{M\ell }}} & {\frac{{2\rho }}{{M\ell }}}  \\
   0 & 0 & 0 & 0 & 0 & 1 & 0 & 0  \\
   {\frac{mg}{M}} & -  \frac{2\lambda}{M}  & 0 & 0 & { - \frac{k}{M}} & { - \frac{{2\rho }}{M}} & {\frac{k}{M}} & 0  \\
   0 & 0 & 0 & 0 & 0 & 0 & 0 & 1  \\
   0 & 0 & {\frac{{mg}}{M}} & -  \frac{2\lambda}{M} & {\frac{k}{M}} & 0 & { - \frac{k}{M}} & { - \frac{{2\rho }}{M}}  
\end{array}} \right)
\left( {\begin{array}{l}
  \theta_1   \\
   v_1   \\
 \theta_2   \\
 v_2  \\ 
 x_1  \\ 
 w_1  \\ 
 x_2  \\ 
 w_2  
\end{array}} \right)\, ,
\label{eq4:2}
\end{equation}
where we have introduced the new variables, $v_1=\dot \theta_1$, $v_2=\dot \theta_2$, $w_1=\dot x_1$, $w_2=\dot x_2$, and the new constants, 
$A=- \omega^2  (1+ m/M)$, and $B=2\lambda  (1/(M\ell)+ 1/(m\ell))$. As, $|\theta_1(t)|<{\tilde \theta}$  and $|\theta_2(t)|<{\tilde \theta}$,  for every $t\in [0,t^*]$, then, for the same initial conditions, the small amplitude solutions of (\ref{eq4:1}) and (\ref{eq4:2}) coincide in the interval $[0,t^*]$. 

 Denoting by $Q$ the matrix in (\ref{eq4:2}), we have, $\hbox{Det}\, Q=0$. A simple inspection of $Q$, shows that $Q$ has only one zero eigenvalue, corresponding to the eigendirection defined by the equation $x_1=x_2$\footnote{The vector $x^T=(0,0,0,0,1,0,1,0)$ is such that $Mx=0$.}.
We investigate now the stability  of the line of fixed points $x_1=x_2$ of both equations (\ref{eq4:1}) and (\ref{eq4:2}). 

\begin{proposition}\label{p2} If $\lambda >0$, $M>0$, $\ell>0$, $m>0$, and $\rho >0$ is sufficiently small, then the systems differential equations (\ref{eq4:1}) and (\ref{eq4:2}) have a line of fixed points with coordinates, $\theta_1=0$, $v_1=0$, $\theta_2=0$, $v_2=0$, $w_1=0$, $w_2=0$, $x_1=x_2$, and this line  of fixed points is Lyapunov unstable. 
\end{proposition}

\begin{proof} We assume that, $\lambda >0$, $M>0$, $\ell>0$ and $m>0$. The existence of the line of fixed points follows by solving the equation $Qy=0$, where,
\[
y^T=(\theta_1,v_1,\theta_2,v_2,x_1,w_1 ,x_2,w_2 )\, .
\] 
For $\rho=0$, the characteristic polynomial of  the matrix $Q$ in (\ref{eq4:2}) is,
\[
q_{\rho=0}(x)=x^2(A+(B-x) x) \left(2 g k m+M
   (A+(B-x) x) \left(M x^2+2 k\right) \ell -4 k x
   \lambda \right)\, ,
\]
where $A=- \omega^2  (1+ m/M)$  and $B=2\lambda  (1/(M\ell)+ 1/(m\ell))$.
As the polynomial $q_{\rho=0}(x)$ has the two roots, 
\[
\lambda=\frac{1}{2}\left(  B\pm \sqrt{B^2+4A} \right)\, ,
\] 
with positive real parts, $B>0$, then, by continuity 
of $Q(\rho)$, for sufficiently small $\rho >0$, the characteristic polynomial $q_{\rho>0}(x)$ of the matrix $Q$,
  has also eigenvalues with positive real parts. Therefore, for sufficiently small $\rho >0$, $\lambda >0$, $M>0$, $\ell>0$, $m>0$, the line of fixed points of the linear system (\ref{eq4:2}) is Lyapunov unstable. As
both systems of equations (\ref{eq4:1}) and (\ref{eq4:2}) have the same phase space orbits near the common line of fixed points, the local properties of the flows are the same, and this is sufficient to prove the local instability of
the flow defined by the system of equations (\ref{eq4:1}).
\qed
\end{proof}

Proposition \ref{p2} gives the conditions of nonconvergence to the quiescent state  of the solutions of the system of linear equation (\ref{eq4:2}). This quiescent state is the line of fixed points $x_1=x_2$ in the eight-dimensional phase space. Under the conditions of the Proposition \ref{p2}, the two pendulum clocks  have sustained oscillations, in the sense that, 
$\lim_{t\to \infty}\theta_1(t)\not=0$, and $\lim_{t\to \infty}\theta_2(t)\not=0$.

To find a sufficient condition 
of existence of anti-phase synchronization of the two non-linear oscillators,
we assume now that the asymptotic solutions of the system of equations (\ref{eq4:1}) synchronize in anti-phase. As the two pendulum clocks are identical, with, $t_n=t_0+n h$, where $n=0,1,\cdots $, and $t_0$ is the initial time, we must have,
\begin{equation}
\begin{array}{lcl}\displaystyle
\displaystyle\lim_{t_n\to \infty}\theta_1(t_n)&=&\displaystyle-\lim_{t_n\to \infty}\theta_2(t_n) \\  \displaystyle
\displaystyle\lim_{t_n\to \infty}x_1(t_n)&=&\displaystyle-\lim_{t_n\to \infty}x_2(t_n) \\  \displaystyle
\displaystyle \lim_{t_n\to \infty}v_1(t_n)&=&\displaystyle-\lim_{t_n\to \infty}v_2(t_n) \\ \displaystyle
\displaystyle \lim_{t_n\to \infty}w_1(t_n)&=&\displaystyle-\lim_{t_n\to \infty}w_2(t_n) \, .
\end{array} 
\label{eq4:3}
\end{equation}
for every $h>0$\footnote{In this context, we say that the two pendulum clocks synchronize in anti-phase if,
for every $h>0$, $\lim_{t_n\to \infty}\theta_1(t_n)=-\lim_{t_n\to \infty}\theta_2(t_n)$. As we shall see in \S~\ref{sec:5}, this definition can only be used in the case of identical pendulum clocks.}.

We now define the new variables,
$\theta  = \theta_1  + \theta_2 $, $v   = v_1  + v_2 $,
$x = x_1  + x_2 $ and $w = w_1  + w_2 $. Then,  adding the corresponding variables in the system of equations (\ref{eq4:2}), we obtain, 
\begin{equation}\displaystyle
\left( {\begin{array}{l}
{\dot \theta }  \\
{\dot v}  \\
{\dot x}  \\
{\dot w}  
\end{array}} \right) = 
\left( {\begin{array}{*{20}c}
0 & 1 & 0 & 0  \\ \displaystyle
 - \omega^2  (1+ \frac{m}{M})\, & 2\lambda  (\frac{1}{M\ell}+ \frac{1}{m\ell})\, & 0 & {\frac{{2\rho
}}{{M\ell }}}  \\ \displaystyle
0 & 0 & 0 & 1  \\ \displaystyle
{\frac{mg}{M}} & -  \frac{2\lambda}{M} & 0 & { - \frac{{2\rho }}{M}}   
\end{array}} \right)
\left( {\begin{array}{*{20}c}
\theta   \\
v  \\
x  \\
w  
\end{array}} \right)\, .
\label{eq4:4}
\end{equation}
In the following, we call the  system of equations (\ref{eq4:4}) the reduced system of equations associated to system (\ref{eq4:2}).
The reduced system of equations (\ref{eq4:4})  has a line of fixed points
with coordinates, $\theta=0$, $v=0$, $w=0$ and 
$x=$constant.
If the line of fixed points of the reduced linear system 
(\ref{eq4:4}) is asymptotically stable, then
we  have,
\begin{equation}
\begin{array}{lclcl}\displaystyle
\lim_{t\to \infty}\theta (t)&=& \displaystyle\lim_{t\to \infty}\theta_1(t)+\lim_{t\to \infty}\theta_2(t) =0\\  \displaystyle
\lim_{t\to \infty}v (t)&=& \displaystyle \lim_{t\to \infty}v_1(t)+\lim_{t\to \infty}v_2(t) =0\\ \displaystyle
\lim_{t\to \infty}w (t)&=& \displaystyle \lim_{t\to \infty}w_1(t)+\lim_{t\to \infty}w_2(t) =0\, .
\end{array} 
\label{eq4:5}
\end{equation}
If, for every $h>0$, the first   condition  in (\ref{eq4:3}) is verified,   the first condition in (\ref{eq4:5}) is also verified, and synchronous solutions of equation (\ref{eq4:1}) are also stable solutions of equations (\ref{eq4:4}).

As, by Proposition~\ref{p2}, $\lim_{t_n\to \infty}\theta_1(t_n)\not=0$, and $\lim_{t_n\to \infty}\theta_2(t_n)\not=0$, for any initial condition away from the line of fixed points, any  solution of the differential
equation (\ref{eq4:1}) that anti-phase synchronizes is also an asymptotically stable solution of the reduced system of equations (\ref{eq4:4}).

Hence, if the line of fixed points of the reduced system 
(\ref{eq4:4}) is asymptotically stable and the line of
fixed points of the linear system of equations (\ref{eq4:2}) is unstable (Proposition~\ref{p2}), then 
the solutions $\theta_1(t)$ and $\theta_2(t)$ of the system of equations (\ref{eq4:1}) anti-phase synchronize. The asymptotic stability condition for the
reduced the linear system (\ref{eq4:4}), together with the instability condition of Proposition \ref{p2}, both give a sufficient condition for the existence of exact anti-phase synchronization of
the two identical non-linear pendulum clocks.   

To analyze the stability properties of the line of fixed points of the system of equations (\ref{eq4:4}),  we calculate the  characteristic polynomial of the  matrix $P$ in (\ref{eq4:4}),
\begin{equation}
\begin{array}{ll}
p(y ) &= y(m M \ell y^3 +y^2(2 m   \ell \rho- 2 m   \lambda - 2 M \lambda  )  \\& +y( m^2   \ell \omega^2 + 
 m M  \ell \omega^2 -  
 4   \lambda \rho)+ 
 2 m  g \rho  )\\
&= yp_1(y)\, .
\end{array}
\label{eq4:7}
\end{equation}
To the eigenvalue $x=0$ corresponds the eigendirection $e_x$.
This eigendirection is the line of fixed points of the linear system (\ref{eq4:4}).  If all the eigenvalues of the polynomial $p_1(y)$ in (\ref{eq4:7})
have negative real parts, any initial condition away from  the line of fixed points $x=$constant,  evolve in time to this line of fixed points. 
Therefore, we have:

\begin{theorem}\label{t1} We consider the system of differential equations (\ref{eq4:1}), with damping function (\ref{eq3:2}). If $\lambda >0$, $\tilde \theta >0$, $m>0$, $\ell >0$, $M>0$, $k>0$,  $g>0$, and $\rho>0$ is sufficiently small, 
and if the reduced linear differential equation (\ref{eq4:4}) has only non-positive eigenvalues, then,   the solutions of equation (\ref{eq4:1}), with damping function (\ref{eq3:2}),  synchronize in anti-phase, in the sense that, for every $h>0$,
\[
\displaystyle \lim_{t_n\to \infty}\theta_1(t_n)=-\lim_{t_n\to \infty}\theta_2(t_n)\, ,
\] 
where, $t_n=t_0+nh$,  $\lim_{t_n\to \infty}\theta_1(t_n)\not= 0$, and $\lim_{t_n\to \infty}\theta_2(t_n)\not= 0$.
Moreover,  
if the polynomial,  
\[
\begin{array}{rl}
q(\rho)=&4 m \ell \lambda \rho^2  - (4m\lambda^2  + 4M\lambda^2  + 
g m^3\ell )\rho  \\&+ g m^3\lambda   
+  2 g m^2 M\lambda +g m M^2\lambda \, ,
\end{array} 
\]
has two real roots $\rho_1$ and $\rho_2$, and 
 $\rho$  obeys to the inequalities,
\[
\begin{array}{lcl}
\rho_1 <&\rho&<\rho_2\\ \displaystyle
&\rho& >  \rho_0=  \frac{\lambda}{\ell} (1 +  \frac{M}{m}) \, ,
\end{array} 
\]
then, for any initial condition away from the line of fixed points of the system of equations (\ref{eq4:1}), the solutions of equation (\ref{eq4:1}), with damping function (\ref{eq3:2}), anti-phase synchronize.
\end{theorem}

\begin{proof} 
The sufficient condition for the existence of anti-phase synchronization of the two pendulum clocks has been derived before the statement of the theorem. The instability of the line of fixed points of the system of equations (\ref{eq4:1}) has been proven in Proposition ~\ref{p2}. To prove the condition of   non-positivity of the eigenvalues of the matrix in the system of equations (\ref{eq4:4}), we  use the Routh-Hurwitz criterion, \cite{10}.
By (\ref{eq4:7}), as $p_1(y)=a_0y^3+a_1y^2+a_2y+a_3$, and
as, by hypothesis, $a_0>0$ and $a_3>0$, by the Routh-Hurwitz criterion, if $a_1>0$ and $(a_1a_2-a_0a_3)>0$, then the polynomial $p_1(y)$ has only roots with negative real parts. As,
\[
\begin{array}{rl}
(a_1a_2-a_0a_3)=-q(\rho)=&-4 m \ell \lambda \rho^2  + (4m\lambda^2  + 4M\lambda^2  + 
g m^3\ell )\rho  \\&- g m^3\lambda   
-  2 g m^2 M\lambda -g m M^2\lambda \, ,
\end{array} 
\]
the polynomial $q(\rho)$ has a global minimum for positive values of $\rho$ and can have two real roots. This proves the first inequality of the theorem. The second inequality follows from the Routh-Hurwitz condition, $a_1=2 m   \ell \rho- 2 m   \lambda - 2 M \lambda  >0$.
\qed
\end{proof}

Theorem \ref{t1} gives a sufficient condition for the existence of exact anti-phase synchronization in the Huygens's two pendulum clocks system. To test numerically the results of Theorem \ref{t1}, we take for the parameters of the  nonlinear oscillator  (\ref{eq3:1})-(\ref{eq3:2})  the  values, $g=9.8$, $m=1$, $\ell=1$, $\lambda =0.1$, and $\tilde \theta=0.1$. 
For these parameter values, the period of the solutions  on the limit cycle of
the equation (\ref{eq3:1}) is, $T=2.008$.
The  parameters describing the interaction between the pendulum clocks have been set to the values, $k=10$, $M=0.1$, and  $\rho$ is a free parameter. 

\begin{figure}
\begin{center}
\includegraphics[width=10.0cm,height=6.38cm]{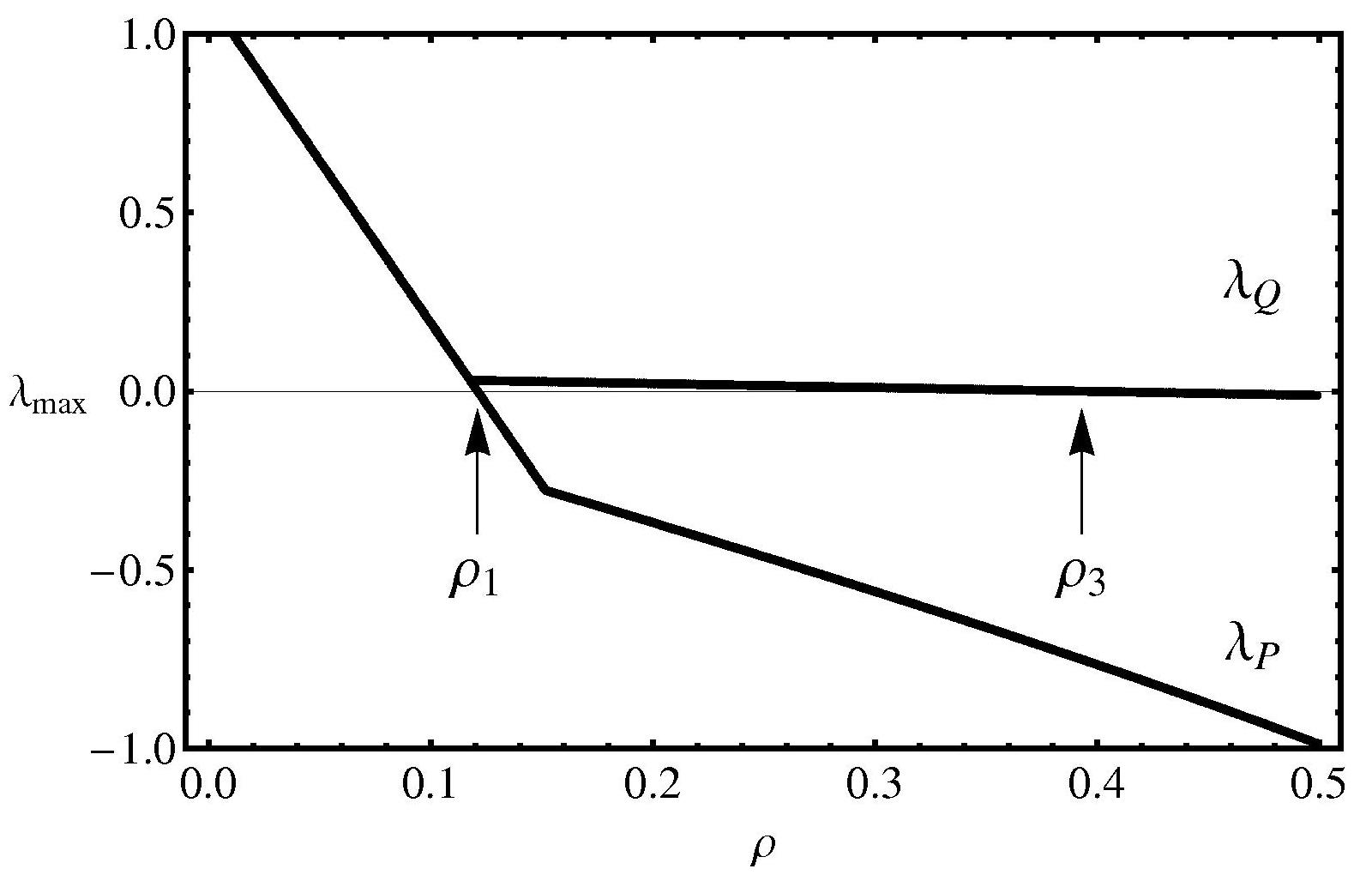}
\end{center}
 \caption{Eigenvalues with
the largest real part of the matrices $Q$ and $P$ of
the linear systems (\ref{eq4:2}) and (\ref{eq4:4}),
respectively, as a function of the damping parameter $\rho$. The other parameter have been fixed to the values: $g=9.8$, $m=1$, $\ell=1$, $\lambda =0.1$,   $\tilde \theta=0.1$, $k=10$ and $M=0.1$. If $\rho>\rho_1=0.121$ and $\rho<\rho_2=24.489$, all the eigenvalues of the matrix $P$ have non-positive real parts, and the line of fixed points of system (\ref{eq4:4}) is Lyapunov stable. If $\rho<\rho_3=0.393$, the matrix $Q$ has
eigenvalues with positive real parts, and the line of fixed points of system (\ref{eq4:1}) is Lyapunov unstable. By Theorem~\ref{t1}, if $\rho \in [\rho_1, \rho_3]$, the two pendulum clocks  synchronize in anti-phase.}
\label{fig3} 
\end{figure}

To test the conditions of Theorem~\ref{t1},
in Fig.~\ref{fig3},
we have plotted the  eigenvalues with
the largest real part of the matrices $Q$ and $P$, of
the linear systems (\ref{eq4:2}) and (\ref{eq4:4}),
respectively,
as a function of the damping parameter $\rho$.  The zero eigenvalue has been excluded from the characteristic polynomials of the matrices $Q$ and $P$. According to Theorem~\ref{t1}, if, $0.121=\rho_1<\rho <\rho_2=24.489$, $\rho>\rho_0=0.11$, and the matrix $Q$ has eigenvalues with positive real parts,  the two pendulum clocks  synchronize in anti-phase. Numerically, for our reference parameter values, if $\rho<\rho_3=0.393$, the matrix $Q$ of the system of equations (\ref{eq4:2}) has
positive eigenvalues. Therefore, the  conditions of Theorem~\ref{t1}
imply that, if $\rho \in [\rho_1, \rho_3]$,  the two pendulum clocks synchronize in anti-phase, Fig.~\ref{fig3}.

\begin{figure}[p]
\begin{center}
\includegraphics[width=7.8cm,height=6.24cm]{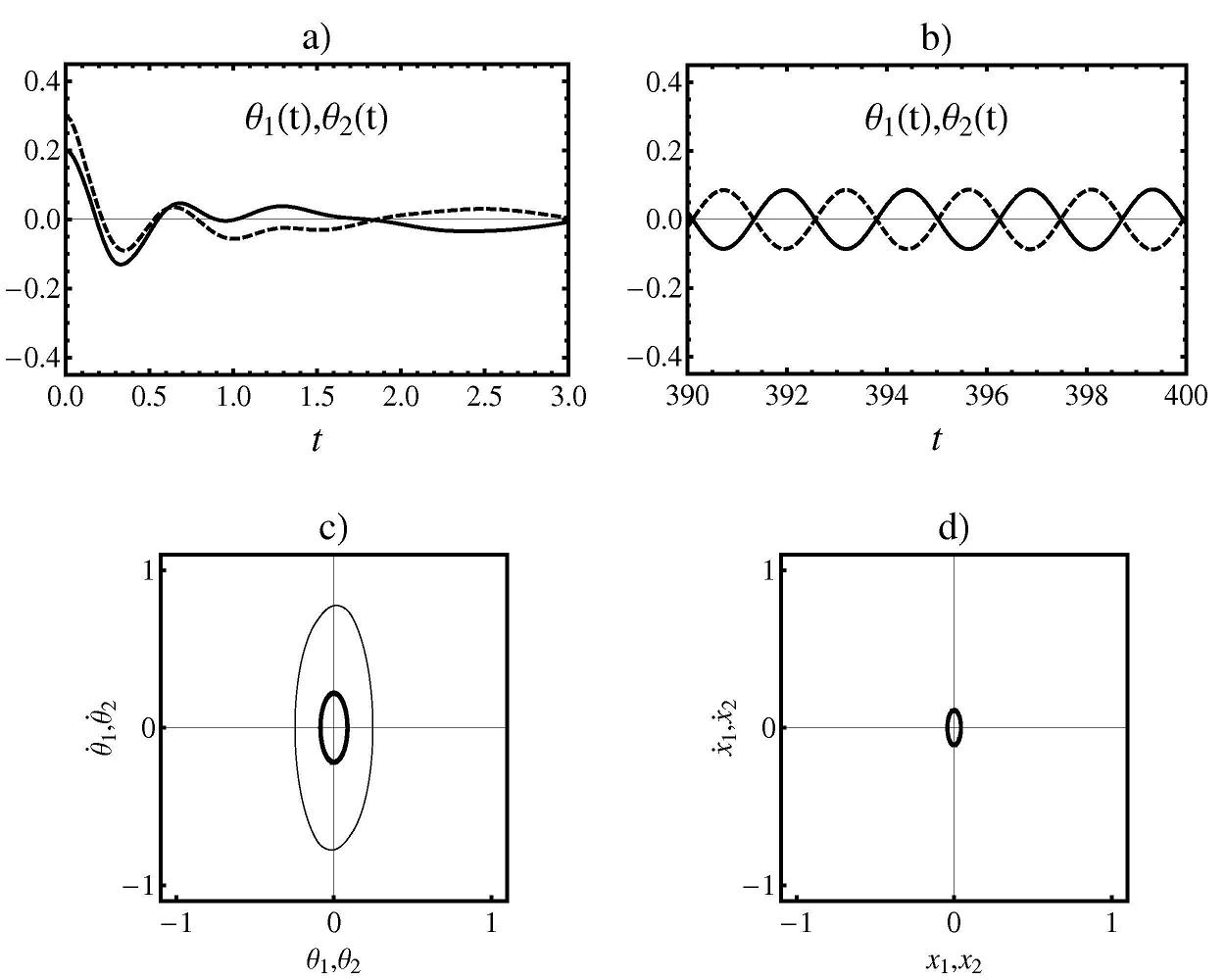}
\end{center}
 \caption{Numerical solutions of the system of equations (\ref{eq4:1}), with damping function (\ref{eq3:2}), describing the coupling of two identical pendulum clocks. The parameter values of the simulation are: $m =
1$,  $\ell  = 1$, $g =9.8$,  $k = 10$, $M =0.1$, $\lambda=0.1$, $\tilde \theta=0.1$ and $\rho  = 0.37$. The initial
conditions are:  $\theta_1(0)=0.2$, $\theta_2(0)=0.3$,
$x_1(0)=0$, $x_2(0)=0$, $\dot \theta_1(0)=0$, $\dot \theta_2(0)=0$, $\dot x_1(0)=0$ and $\dot x_2(0)=0$. In a) and b), we show the time evolution of the angular coordinates of the two pendulum clocks, before and after anti-phase synchronization, respectively.
In c) and d), we show the asymptotic solutions in the reduced phase space of the two pendulum clocks  (c), and of the two attachment points  (d). For comparison, in c), we show the limit cycle solution (thin line curve) of the reference equation (\ref{eq3:1}), with damping function (\ref{eq3:2}).The period of the anti-phase oscillations is $T=2.457$. }
\label{fig4} 
\end{figure}

By  Theorem \ref{t1}, the anti-phase synchronization between the two non-linear pendulum oscillators exists  for  $\rho\in [\rho_1=0.121, \rho_3=0.393]$. 
In Fig.~\ref{fig4}, we show the time evolution of the angular coordinates and attachment points of the two pendulum clocks, for $\rho  = 0.37$, and calculated numerically from the system of equation (\ref{eq4:1}). We have chosen for initial conditions the coordinate values, $\theta_1(0)=0.2$, $\theta_2(0)=0.3$,
$x_1(0)=0$, $x_2(0)=0$, $\dot \theta_1(0)=0$, $\dot \theta_2(0)=0$, $\dot x_1(0)=0$ and $\dot x_2(0)=0$. The two non-linear pendulums and the attachment points  synchronize in anti-phase. Numerically, the period of the synchronized state is $T=2.457$, which differs from the eigen-period of the uncoupled non-linear oscillators, $T=2.008$. The anti-phase synchronized state of the system of equations (\ref{eq4:1}) corresponds to a   stable limit cycle in the eight-dimensional phase space.

\begin{figure}
\begin{center}
\includegraphics[width=7.8cm,height=6.20cm]{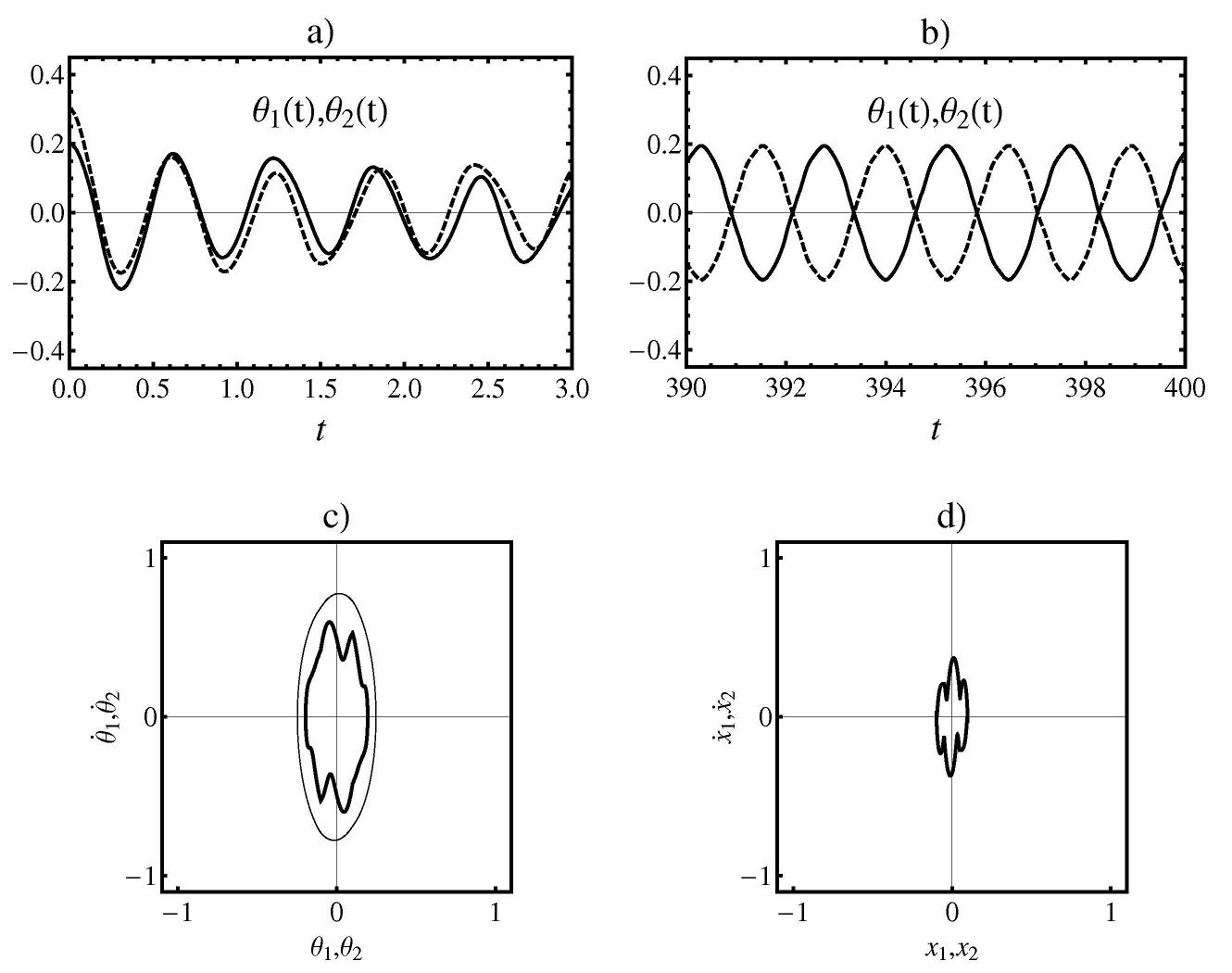}
\end{center}
 \caption{Anti-phase synchronization of the two pendulum clock system, for the same parameter values of Fig.~\ref{fig4}, except for the damping parameter $\rho$ that, in this case, has the value $\rho=0.1$. In this simulation,  the period of the exact anti-phase oscillations is $T=2.463$. In c), the thin line curve is the limit cycle solution of the reference equation (\ref{eq3:1}), with damping function (\ref{eq3:2}).}
\label{fig5} 
\end{figure}

\begin{figure}[p]
\begin{center}
\includegraphics[width=8cm,height=6.55cm]{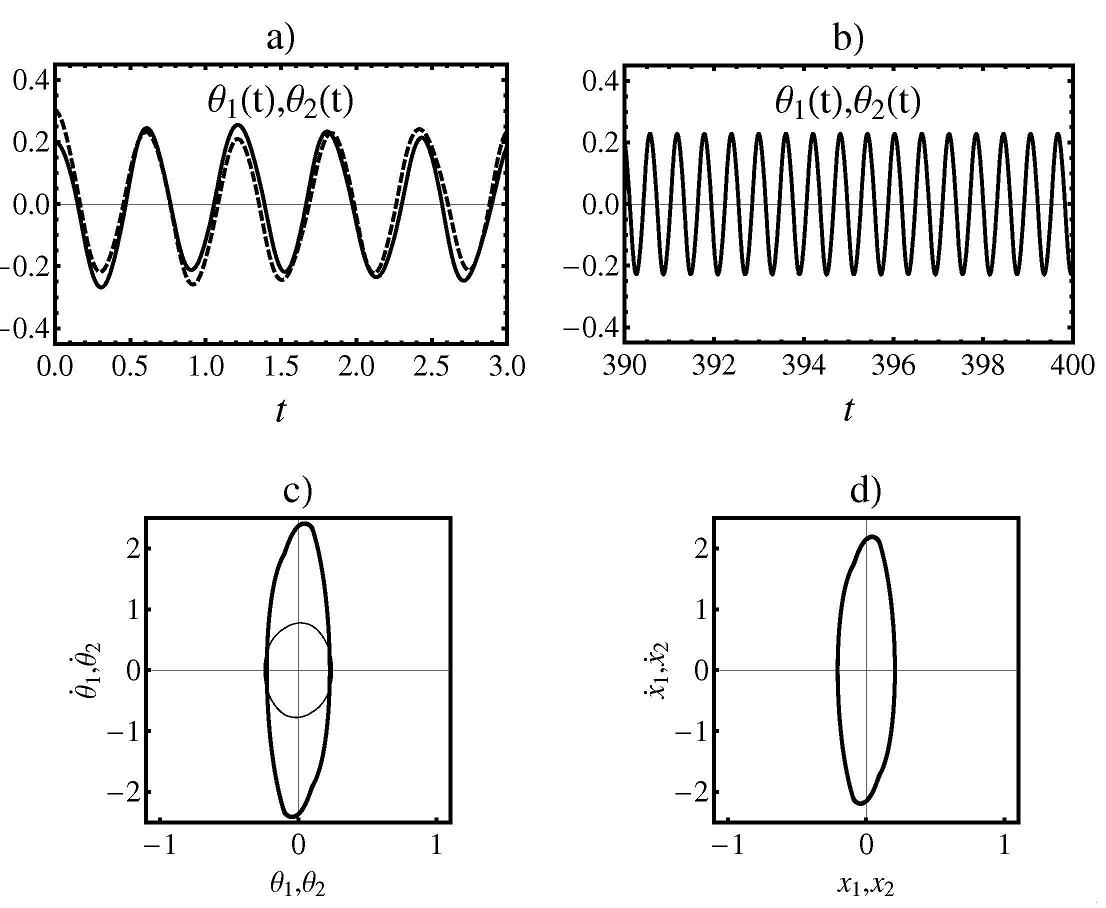}
\end{center}
 \caption{In-phase synchronization of the two pendulum clock system, for the same parameter values of Fig.~\ref{fig4}, except for the damping parameter $\rho$ that, in this case, has the value $\rho=0.01$. In this simulation,  the initial conditions are :  $\theta_1(0)=0.2$, $\theta_2(0)=0.3$,
$x_1(0)=0$, $x_2(0)=0$, $\dot \theta_1(0)=0$, $\dot \theta_2(0)=0$, $\dot x_1(0)=0$ and $\dot x_2(0)=0$. The period of the in-phase oscillations is $T=0.606$. In c), the thin line curve is the limit cycle solution of the reference equation (\ref{eq3:1}), with damping function (\ref{eq3:2}).}
\label{fig6} 
\end{figure}

\begin{figure}
\begin{center}
\includegraphics[width=8cm,height=6.54cm]{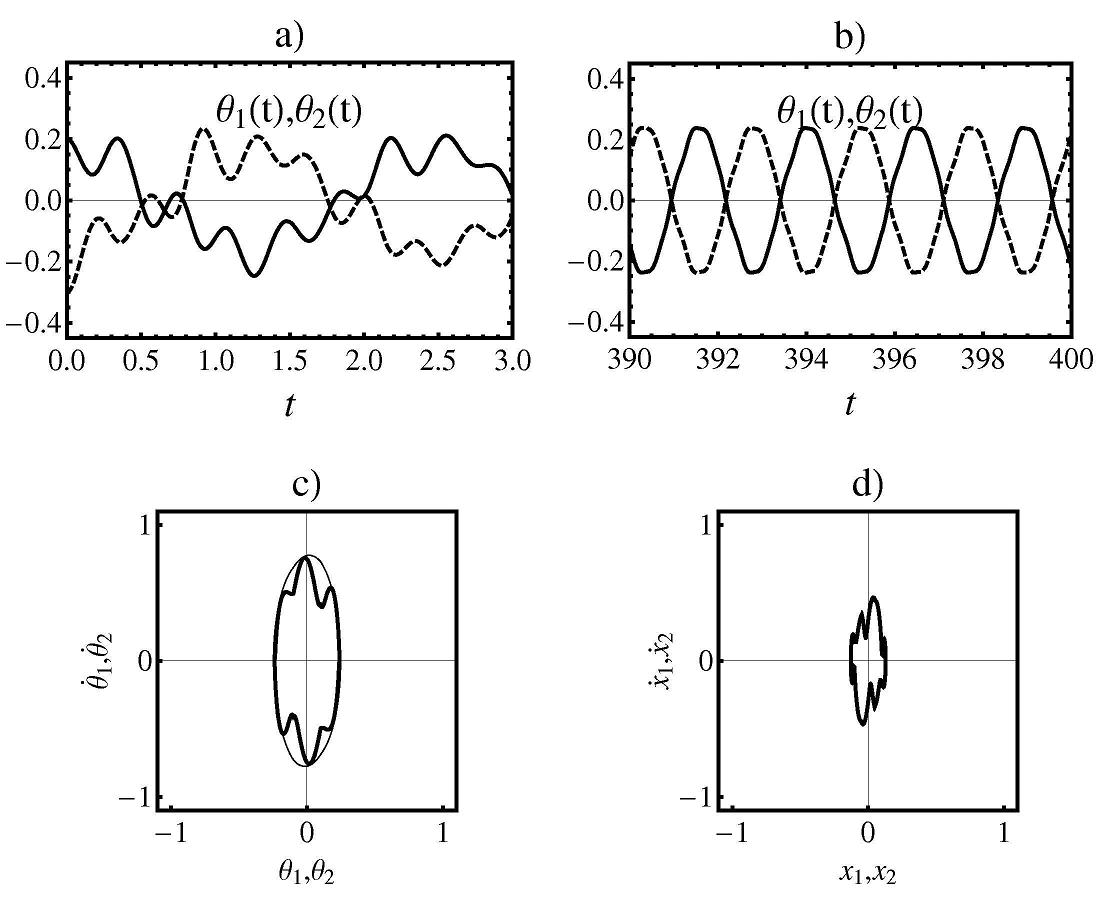}
\end{center}
 \caption{Anti-phase synchronization of the two pendulum clock system, for the same parameter values of Fig.~\ref{fig6}, and $\rho=0.01$. In this simulation,  the initial conditions are :  $\theta_1(0)=0.2$, $\theta_2(0)=-0.3$,
$x_1(0)=0$, $x_2(0)=0$, $\dot \theta_1(0)=0$, $\dot \theta_2(0)=0$, $\dot x_1(0)=0$ and $\dot x_2(0)=0$. Initially, the two pendulums are approximately in anti-phase.
The period of the anti-phase oscillations is $T=2.463$. In c), the thin line curve is the limit cycle solution of the reference equation (\ref{eq3:1}), with damping function (\ref{eq3:2}).}
\label{fig7} 
\end{figure}

\begin{figure}
\begin{center}
\includegraphics[width=9cm,height=9.19cm]{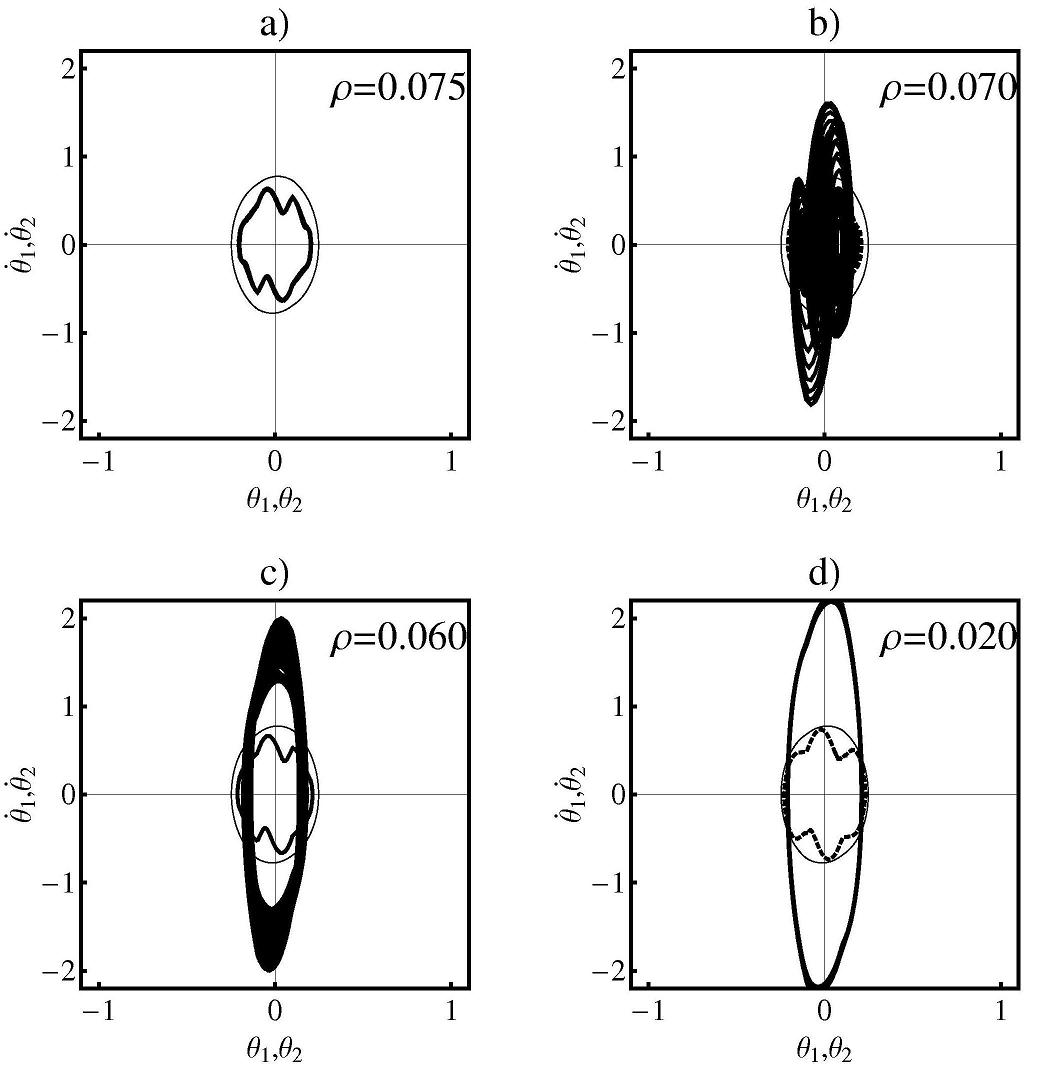}
\end{center}
 \caption{Coexistence of anti-phase and in-phase synchronization and aperiodic regimes for the asymptotic solutions of the system of equations (\ref{eq4:1}). We show the limit cycle solutions of the system equations (\ref{eq4:1}) for the same parameter values of Fig.~\ref{fig4}, and several values of $\rho$: a) $\rho=0.075$, b) $\rho=0.07$, c) $\rho=0.06$, and d) $\rho=0.02$. We have plotted the asymptotic solutions in phase space for two different initial conditions. In one case we have taken $\theta_1(0)=0.2$ and $\theta_1(0)=0.3$, with the other initial conditions equal to zero. In the other case, we have taken  $\theta_1(0)=0.2$ and $\theta_1(0)=-0.3$. For $\rho> 0.075$, different initial conditions leads always to anti-phase synchronization. For $\rho\le 0.02$, we have  two limit cycles in phase space, one corresponding to anti-phase synchronization, and the other to  in-phase synchronization. The thin line curve is the limit cycle solution of the reference equation (\ref{eq3:1}), with damping function (\ref{eq3:2}).}
\label{fig8} 
\end{figure}

In Fig.~\ref{fig5}, we have decreased the parameter $\rho$ to values below the transition values $\rho_0$ and $\rho_1$ of Theorem \ref{t1}. In this case, the two non-linear oscillators  also anti-phase synchronize. The numerical integration for several initial conditions shows that, for the damping parameter value $\rho=0.1$,  the system of equation (\ref{eq4:1}) has a stable limit cycle in the eight-dimensional phase space. In this case, we are outside the conditions of Theorem~\ref{t1}. 

Decreasing further the parameter $\rho$, for $\rho<0.06$, and for the same initial conditions as in Fig.\ref{fig4}, the two oscillators synchronize with the same phase (in-phase), Fig.~\ref{fig6}. However, changing the initial conditions for an approximate anti-phase initial state of the two pendulums, we obtain anti-phase synchronization, Fig.~\ref{fig7}.  This simple fact shows that, for $\rho<0.07$, there are two stable limit cycles in the eight-dimensional phase space, and these limit cycles have their own basins of attraction.

To analyze the transition from the anti-phase to the in-phase synchronization asymptotic regimes, we have chosen $\rho$ in the interval $[0.06,0.07]$, and we have changed the initial conditions of the numerical simulations. 
For $\rho$ in the interval $[0.06,0.07]$, the initial conditions that eventually lead to an in-phase synchronized regime show quasi-periodic behavior in time.
If $\rho<0.06$, the asymptotic  in-phase synchronized state is a limit cycle in phase space. This suggests
the existence of a non-local bifurcation for $\rho$ near the   value $\rho\simeq 0.07$. 

In Fig.~\ref{fig8}, we show the transition from the anti-phase  to the in-phase synchronized asymptotic state  as well as the bifurcation behavior of the attractors. We have superimposed in the same figure the two stable limit cycles that can be reached from different initial conditions. In Fig.~\ref{fig8}a), $\rho=0.075$, there is only one asymptotic anti-phase synchronized state. In Fig.~\ref{fig8}d), $\rho=0.020$, there are two stable asymptotic synchronized state, one in-phase and the other in anti-phase. In Fig.~\ref{fig8}b) and \ref{fig8}c), we have a stable limit cycle for the asymptotic anti-phase synchronized state, but the asymptotic in-phase synchronized sate  
shows quasi-period behavior.

Therefore, we have shown that the interaction mechanism
proposed in section \ref{sec:2} leads asymptotically in time to exact anti-phase synchronization of
coupled identical oscillators. For the same parameter
values, we can have anti-phase and in-phase synchronization, determined by different initial conditions, and the 
tuning between the two types of synchronization regimes can also be controlled through the  damping constant $\rho$.
Further numerical tests show that these results are still
true in the singular limit $M\to 0$.

\section{Synchronization of two pendulum clocks  with different parameters: Robustness}
\label{sec:5}

In the previous analysis, we have consider that both oscillators are characterized by the same parameters. However, in real experiments this is not realistic and we must consider the persistence 
of synchronization when the  parameters of the pendulums are different. Here, the persistence of the anti-phase and the in-phase synchronization states is analyzed numerically. 
We take, $m_1=m$, $m_2  = m (1 + \varepsilon
)$, $\ell_1=\ell$ and $\ell_2  = \ell (1 + \delta )$,
where $\varepsilon$ and $\delta$ can have positive or negative values. In this case,
 the equations of motion (\ref{eq6}) are rewritten as,
\begin{equation}
\begin{array}{l}\displaystyle
\ddot \theta_1  + (\frac{1}{m\ell}+\frac{1}{M\ell})f (\theta_1 ;\lambda ,\tilde \theta )\dot \theta_1 + \omega^2(1+\frac{m}{M})\theta_1  -2\frac{\rho}{M\ell}\dot x_1=  - \frac{k}{M\ell} (x_2  - x_1 )  \\[6pt] \displaystyle
\ddot \theta_2  + \frac{1}{ (1 + \varepsilon
)(1 + \delta )}(\frac{1}{m\ell}+\frac{1}{M\ell})f (\theta_2 ;\lambda ,\tilde \theta )\dot \theta_2 + \frac{\omega^2}{(1 + \delta )}(1+\frac{m}{M}(1 + \varepsilon
))\theta_2  \\[6pt] \displaystyle
-2\frac{\rho}{M\ell (1 + \delta )}\dot x_2=   \frac{k}{M\ell (1 + \delta )} (x_2  - x_1 )\\[6pt] \displaystyle
\ddot x_1  - \frac{1}{M}  f (\theta_1 ;\lambda ,\tilde \theta )\dot \theta_1 -\frac{m}{M} 
g \theta_1   + 2\frac{\rho}{M} \dot x_1   = \frac{k}{M}(x_2  - x_1 )    \\[6pt]   \displaystyle
\ddot x_2 - \frac{1}{M}  f (\theta_2 ;\lambda ,\tilde \theta )\dot \theta_2 -\frac{m}{M} (1 + \varepsilon
)
g \theta_2  + 2\frac{\rho}{M} \dot x_2   = -\frac{k}{M}(x_2  - x_1 )\, , 
\end{array}
\label{eq5:1}
\end{equation}
where $f(\theta; \lambda ,{\tilde \theta})$ is given by (\ref{eq3:2}).

\begin{figure}
\begin{center}
\includegraphics[width=9cm,height=7.08cm]{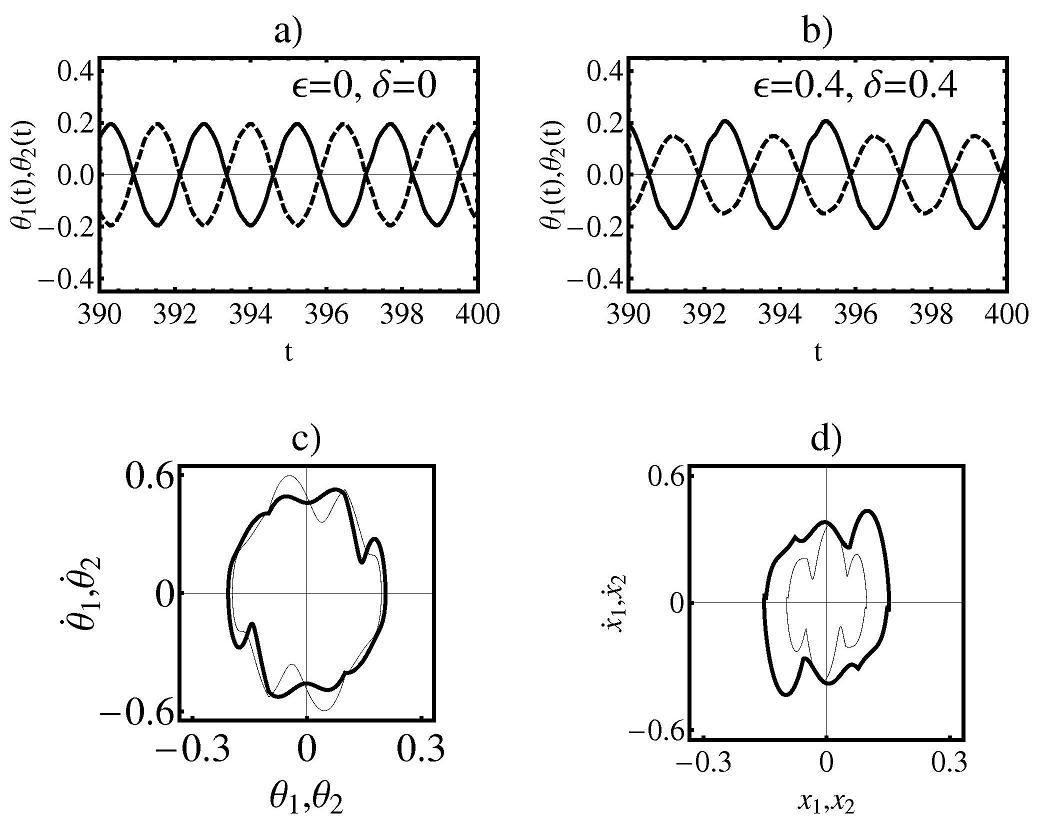}
\end{center}
 \caption{Anti-phase synchronization of two pendulum clocks with different lengths and masses. In a), we have the anti-phase synchronized state as in Fig.~\ref{fig5}. In b), we show the anti-phase synchronized state for $\varepsilon=0.4$ and $\delta=0.4$. In c) and d), we show the limit cycles in phase space (thick lines) of the angular coordinates and of the attachment points of the two pendulums. The thin lines are the limit cycles for the cases $\varepsilon=0$ and $\delta=0$. In b), the period of oscillation is $T=2.661$, and  in   a), $T=2.463$. When the two non-linear pendulums have different parameters, the anti-phase synchronization condition,  $\lim_{t_n\to \infty} \theta_1(t_n)= -\lim_{t_n\to \infty} \theta_2(t_n)$, derived in \S~\ref{sec:4},  is no longer verified.}
\label{fig9} 
\end{figure}

We have integrated numerically the system of equations (\ref{eq5:1}) for the same parameter values of Fig.~\ref{fig5}, and the same initial conditions, but with $\varepsilon=0.4$ and $\delta=0.4$. The numerical results are presented in Fig.~\ref{fig9}, and, we  conclude that the anti-phase synchronized state still exists for large values of the parameters $\varepsilon$ and $\delta$.

Fixing $\varepsilon$ to the value $\varepsilon=0.4$,   the anti-phase synchronized state still persists for $\delta\in [-0.1,3.5]$. For $\delta=-0.2$, and the same initial conditions as in Fig.~\ref{fig9}, the system in-phase synchronizes. 

The comparison between figures Fig.~\ref{fig9}a) and
Fig.~\ref{fig9}b) shows that the definition of anti-phase  synchronization used in the previous section is specific to the case of non-linear oscillators with equal parameters.  In Fig.~\ref{fig9}b), the two pendulums clearly synchronize in anti-phase, but, $\lim_{t_n\to \infty} \theta_1(t_n)\not= -\lim_{t_n\to \infty} \theta_2(t_n)$.

As the  model presented here shows anti-phase and in-phase synchronization phenomena for oscillators characterized by different parameters, 
these numerical results show  that the equality between the eigen-periods of the two pendulums is not required to obtain synchronization.

\section{Conclusions}
\label{sec:6}

We have proposed a model of interaction between oscillators leading to  exact  anti-phase  synchronization.  This phenomena has been observed by the first time, in 1665,   by Christiaan Huygens. 

The interactions parameters in our model are a damping constant $\rho$, a stiffness constant $k$ of a linear string, and a mass parameter $M$. This mass parameter is associated with the interaction, and not with the individual oscillators. For moderated values of the (wet) damping constant $\rho$,  $k>0$, and $M>0$, the asymptotic solutions of the model equations converge to a stable limit cycle in the eight-dimensional phase space of the model equations. Numerically, this limit cycle is the only stable attractor of the dynamics of the interacting oscillators.

For smaller values of the damping constant $\rho$, two stable limit cycles in phase space coexist. One corresponds to the anti-phase synchronized state of the two pendulum clocks, and the other corresponds to the in-phase synchronized state. The two limits cycles are reached by different initial conditions in the model equations.

The transition between the one-limit cycle solution and the two-limit cycle solution appears by a global bifurcation tuned by $\rho$. After the bifurcation, 
the anti-phase state coexists with an approximate quasi-periodic in-phase state. 

Changing the parameters of the individual pendulum clocks, we obtain  the same synchronization 
properties as in the case of oscillators with identical parameters. This fact shows that the interaction mechanism purposed here is robust to changes in the parameters of the nonlinear oscillators.

In all the cases analyzed numerically, the anti-phase and the in-phase synchrony occurs  with periods
different from the eigen-periods of the individual oscillators. This shows that the equality between the eigen-periods of the individual oscillators is not required to obtain anti-phase or in-phase synchronization.

An important new issue introduced in the model is the possibility of existence of small movements of the attachment points of the pendulums clocks, a situation
clearly avoid in the modern experimental devices, and
diverging from mechanism of synchrony proposed by Kortweg, \cite{kort}.
This explains why 
modern experimental setups have not been able to
reproduce the original Huygens's results.

\begin{acknowledgement}
I would like to thank the support of the Ettore Majorana Center for Scientific Culture, and the 
hospitality of the organizers of the conference "Variational Analysis and Aerospace Engineering", dedicated to Prof. Angelo Miele for his 85th birthday. 
This work has been partially supported by a Funda\c c\~ao para a Ci\^encia e a Tecnologia (FCT) pluriannual funding grant to the NonLinear Dynamics Group (GDNL).
\end{acknowledgement}

\end{document}